# How much does an interlibrary loan request cost? A review of the literature

Marc-André Simard[1], Jason Priem[2], and Heather Piwowar[2]
[1]École de bibliothéconomie et des sciences de l'information, Université de Montréal, Québec, Canada.
[2]Our Research(https://ourresearch.org/)
Contact: marc-andre.simard.1@umontreal.ca

## Abstract

Interlibrary loan (ILL) services are used to fill the gap between academic libraries' collections and the information needs of their users. Today's trend toward the cancellation of serials "Big Deals" has increased the importance of clear information around ILL to support decision-making. In order to plan the cancellation of a journal package, academic libraries need to be able to forecast their total spendings on ILL, which requires to have an appropriate estimate of what it costs to fulfill an individual ILL request. This paper aims to help librarians answer this question by reviewing the most recent academic literature related to these costs. There are several factors that may affect the cost of an ILL service, including the cost of labour, the geographic location of the library, the use of a ILL software, and membership to a library consortium. We find that there is a wide range of estimates for ILL cost, from $3.75 (USD) to $100.00 (USD). However, Jackson's (2004) figure of $17.50 (USD) per transaction remains the guideline for most researchers and librarians.

## Introduction

Recent studies have shown that cancelling a subscription to a major scientific publishers (also known as "Big Deals'") may lead to massive savings while limiting the impact on the access to the scientific literature that is actually used by the users (i.e Pedersen et al., 2014; Nabe and Fowler, 2015). Several universities, such as Cornell University, MIT, and the University of California, have already cancelled their big deals. SPARC is currently hosting an exhaustive list of institutions who cancelled their subscriptions to major scientific publishers (https://sparcopen.org/our-work/big-deal-cancellation-tracking/). One of the most common ways to assess the impact of Big Deals cancellations is using cost-per-use data. However, when a library makes the decision to cancel journal subscriptions, they also have to take into consideration the impact it may have on the other services they offer. One of the main concerns expressed over the cancellations of big deals is the effect that they could have on the interlibrary loan (ILL) services. ILL are generally used to fill the gap between academic libraries' collections and what their users actually need. Logically, since cancelling Big Deals leads to removing access to scientific resources, it is generally expected that the number of ILLs would increase overtime which would, in turn, lead to an increase in ILL borrowing costs for libraries. In other words, academic libraries that plan to unsubscribe to Big Deals need to be able to forecast their total spendings on ILL. In order to do that, they need to have an appropriate estimate of what it costs to fulfill an individual ILL request. This paper aims to help librarians answer this question by reviewing the most recent scientific evidence related to these costs.

# How much do ILL requests really cost?

When calculating ILL costs, there are several things that need to be taken into consideration. Most studies calculate ILL costs by dividing every cost associated with ILL (including the costs of labour, borrowing and lending) by the number of ILL transactions over a specific period of time. On average, labour costs associated with ILL can account between 36% to 80% of total ILL costs (Chan, 2004; Jackson, 2004; Morris, 2004). In addition, ILL costs are often separated into three categories: 1) borrowing costs, which is the price paid by the borrowing library, 2) lending costs, which is the price paid by the lending library, and 3) the total costs combining borrowing and lending costs. Since this study aims to assist librarians in assessing the effect of Big Deal cancellations, it will only focus on ILL borrowing costs. Table 1 summarizes a list of recent studies that have explored ILL spending.

Looking at borrowing costs, we found data in 15 studies, including 14 at the national level and one at the international level. Reported ILL borrowing costs vary from $3.75 in UK health libraries to as high as $100.00 for international ILL borrowing in Academic Libraries in the U.S. Pacific Northwest. In Australia, the National Library reported in 2001 an average ILL borrowing cost of $32.10 (USD) per transaction, whereas Chan (2004) reported ILL borrowing costs of $33.00 in Hong Kong.

To this day, Jackson's (2004) report for the Association of Research Libraries (ARL) remains the most commonly used figure ($17.50 for borrowing) in recent literature. It offers the most exhaustive analyses using the biggest sample (n = 440,827) out of all the studies featured in this paper. According to Google Scholar (GS), it also has received the highest number citations (n = 97) among all ILL themed studies and has been cited by the vast majority of the studies we found. To our knowledge, there has not been a study as comprehensive on ILL costs since then, with the possible exception of Leon and Kress (2012).

Some libraries choose to split the costs of ILL fees with its users (National Library of Australia, 2001; Tahim et al., 2012) with an average of $10.28 (USD) charged to patrons, while others pay the entire fees. Brown (2012) reported that 5 to 10% of the copyright clearance fees were charged to affiliated or unaffiliated patrons. According to the National Library of Australia (2001), the median price in fees patrons were willing to pay for ILL requests was $0.00 with an average of $4.09 (USD).

Some studies have also made the distinction in ILL borrowing costs between documents and copies of documents. Based on a sample of 23 medium to large academic libraries in the United States, Leon and Kress (2012) reported an average of $7.93 for borrowing copies and $12.11 for borrowing documents, a difference of $4.18. Using a sample of three universities, the University of Oregon, the University of Washington, and the Washington State University in Vancouver, Bean et al. (2012) found an average costs of international ILL borrowing costs of $100.00 for documents with a six to eight weeks delay compared to a $80.00 borrowing cost for photocopies with a four to six weeks delay.

In certain cases, copyright clearance fees may also need to be considered when borrowing or lending a scientific paper as stated by Brown (2012):

Approximately 70% of the McGoogan Library's borrowing article requests are for materials published within the last 5 years and are, therefore, not covered by the fair use provisions of section 108 of the US Copyright Law, as interpreted by the Commission on New Technological Uses of Copyrighted Works (CONTU). The CONTU guidelines, which further clarify subsection 108(g)(2), state that fair use allows libraries to obtain up to 5 articles dated within the past 5 years from a single journal title through ILL on a yearly basis. This guideline is often referred to as the ''rule of five.'' The sixth and subsequently requested articles from a particular title require permission from the copyright holder.

According to different studies, these copyright clearances fees can go as high as $40.00 per article (Reighart & Oberlander, 2008; Blecic et al., 2013; Brown, 2012; England and Jones, 2014), which can be more expensive than buying the article itself. Out of the 15 studies found, four of them mentioned the obligation of paying for copyright clearance fees when using ILL for scientific papers. In all three studies, the average copyright fees paid by the sampled libraries were between $9.00 and $40.00. In most cases the fees are almost entirely absorbed by the library, with the exception of Brown (2012) as previously stated.

It is also possible to lower ILL costs by being a part of an efficient local partnership with other academic libraries such as the Shared Library Service Platform in Québec (LSP; https://www.mcgill.ca/library/about/shared-platform). For instance, Jackson (2004) showed that library consortiums using RapidILL software managed to lower their ILL borrowing costs to $5.21, an economy of over $12.00 per transaction. Leon and Kress (2012) also reported similar savings ($8.23) with an average ILL borrowing cost of $3.85 ($4.70 lending) when being a part of a library consortium.

Table 1. Average interlibrary loan expenditures according to recent studies with their number of citations on Google Scholar. All costs are in USD.

| Study | Borrowing costs | Numbers of transactions | Number of citations on Google Scholar |
| --- | --- | --- | --- |
| Naylor (1997) | $8.51; $4.68 for unfilled transactions | 13,088 | 13 |
| National Library of Australia (2001) | $32.10 | 440,827 | 14 |
| Chan (2004) | $33.00 | 345 titles | 30 |
| Jackson (2004) | $17.50 | 822,384 | 97 |
| Reighart & Oberlander (2008) | - | - | 27 |
| Tahim et al. (2012) | $3.75 (average) | 110 libraries (survey) | 7 |

| Bean et al. (2012) | $100.00 for documents; $80.00 for copies for international ILL | 1,101 | 4 |
| --- | --- | --- | --- |
| Brown (2012) | $11.00 (estimated) | - | 29 |
| Leon & Kress (2012) | $12.11 for documents; $7.93 for copies | 13,875 (mean based on 5 libraries) | 41 |
| McGrath (2012) | $7.35 per ILL through RLUK; $9.02 for ILL through BLDSC | - | 13 |
| Blecic et al. (2013) | $12.00 | - | 47 |
| England & Jones (2014) | $5.21 | - | 18 |
| Lemley & Li (2015) | $15.35 | - | 19 |
| Stefany et al. (2015) | Between $8.00 and $10.00 | 26,079 | 0 |
| Shrauger & Sharf (2017) | $15.00 (estimation) | 9,255 | 4 |

## Conclusion

At first glance, the literature presents a disconcertingly wide range of answers to the question, "how much does an ILL request cost?" When it comes to supporting decision-making, this can leave librarians in a difficult situation. How can they meaningfully budget for ILL when estimates of its cost vary by a factor of thirty, from $3.75 to $100?

However, a closer look at the extant literature shows that there is more room for confidence than a quick overview might indicate. The highest estimate for same-country ILL (the vast majority of use-cases for ILL) is only $32.10, as found by the National Library of Australia. And the lowest estimate of $3.75 (Tahim et al., 2012) may be partly explained by its survey methodology, which few other studies have employed.

For the time being, we recommend use of Jackson's (2004) figure, particularly because it examines by the largest sample of transactions and does so with methodological rigor. These features, combined with its being commissioned by a prominent library organization (ARL), probably explain why it remains the most cited study in the field, by a wide margin. One possible adjustment to Jackson's figure would be to account for inflation; based on the U.S Bureau of Labor Statistics Consumer Price Index (CPI) calculator (https://www.bls.gov/data/inflation_calculator.htm), $17.50 in 2004 would be worth $20.48 today. However, there have been questions raised as to the suitability of the CPI in adjusting for inflation in the scholarly communications system. (Davis, 2009)

There are several factors that may affect the cost of an ILL service, including the cost of labour (i.e. Jackson, 2004), the geographic location of the library (Chen, 2004; Bean et al., 2012; McGrath, 2012), the use of ILL software (Jackson, 2004), and membership to a library consortium (Jackson, 2004; Leon & Kress, 2012). We encourage further research into these in order to help libraries fine-tune estimates for their particular situation. There is no one-size-fits-all answer to the question of ILL cost, and estimates should reflect this.

## Acknowledgement

The authors of this paper would like to acknowledge *Our Research* for the funding and support.

## Author order

The second and third authors contributed equally, and their order was determined by coin flip